# Superionic silica-water and silica-hydrogen compounds under high pressure

Hao Gao[1], Cong Liu[1], Jiuyang Shi[1], Shuning Pan[1], Tianheng Huang[1], Xiancai Lu[2], Hui-Tian Wang[1], Dingyu Xing[1], and Jian Sun[1,*]

[1] National Laboratory of Solid State Microstructures, School of Physics and Collaborative Innovation Center of Advanced Microstructures, Nanjing University, Nanjing 210093, China

[2] State Key Laboratory for Mineral Deposits Research, School of Earth Sciences and Engineering, Nanjing University, Nanjing 210093, China.

Silica, water and hydrogen are known to be the major components of celestial bodies, and have significant influence on the formation and evolution of giant planets, such as Uranus and Neptune. Thus, it is of fundamental importance to investigate their states and possible reactions under the planetary conditions. Here, using advanced crystal structure searches and first-principles calculations in the Si-O-H system, we find that a silica-water compound $(SiO_2)_2(H2O)$ and a silica-hydrogen compound $SiO_2H_2$ can exist under high pressures above 450 and 650 GPa, respectively. Further simulations reveal that, at high pressure and high temperature conditions corresponding to the interiors of Uranus and Neptune, these compounds exhibit superionic behavior, in which protons diffuse freely like liquid while the silicon and oxygen framework is fixed as solid. Therefore, these superionic silica-water and silica-hydrogen compounds could be regarded as important components of the deep mantle or core of giants, which also provides an alternative origin for their anomalous magnetic fields. These unexpected physical and chemical properties of the most common natural materials at high pressure offer key clues to understand some abstruse issues including demixing and erosion of the core in giant planets, and shed light on building reliable models for solar giants and exoplanets.

---

[*] Corresponding author. jiansun@nju.edu.cn

Materials under high pressure and high temperature is of fundamental importance in condensed matter physics, chemistry and planetary sciences, which plays important roles in the structure, compositions and evolution of giant planets including Uranus and Neptune [1–4]. Whether Uranus and Neptune are "icy" giants or "rocky" giants is still a longstanding open question, and the states of their compositions are the key missing parts for solving this mysterious puzzle about the interiors of planets [5,6]. In the current common knowledge, these planets are usually considered to be "icy" because their mantles are believed to be mainly composed by icy mixture of water, ammonia and methane [6–8]. The icy models are mainly supported by the following facts. Firstly, both Uranus and Neptune are far from Sun so that the low temperature leads to the formation of ice. Secondly, the unusual magnetic fields of Uranus and Neptune indicate that there should be a convective and conducting shell around a silicate core and the electrical conductivity is considered to originate from superionic ice phases [1,3,9–12]. Moreover, the tropospheric CO profiles in Uranus and Neptune suggest high O/H enrichments, which is consistent with large fractions of ice in the icy giant models [6,13].

However, current observations of Uranus and Neptune are still limited and it is reported that both ice- and rock-dominated interior models can explain most of the observed fundamental physical properties [14], which means that whether Uranus and Neptune are icy or rocky planets is still unclear [5,15]. Except for the advantages of icy giant models discussed above, its main difficulty is that the O/H enrichments inferred from the CO observations and D/H ratios are inconsistent [6,14], and currently there is no satisfying explanation for this problem. On the other hand, rocky planet models provide a simple way to explain the formation of Neptune, and the lower O/H enrichment is also compatible to the C/H and D/H ratios [14]. A remaining issue of rock interior models is how to explain the magnetic dynamos, which means that conductive materials other than the superionic ice/ammonia are required. Along this line, Soubiran *et al*. [16] found that the electrical conductivity of liquid silicates at high pressures and temperatures is high enough to generate magnetic fields of Super-Earths. Shock experiments on silica indicate that it becomes conductive at more than 500 GPa [2], however, the corresponding temperature ($>10^4$ K) is too high for Uranus and Neptune. Therefore, the conductive materials responsible for the magnetic fields in the solar giants are still lacked in the rock-dominated models [14].

Another key open question about these planets is whether the transitional zone between different layers is smooth or separated [4,6]. In the current three-layer models, there are two boundaries: one is between the H/He atmosphere and the interior while the other is between the icy mantle and the rocky core. For the former boundary, recent theoretical works have proposed reactions between H/He and mantle compositions such as water, ammonia and methane [17–24]. These compounds might exist at the atmosphere-interior boundaries on Uranus and Neptune. However, discussions on possible reactions at the core-mantle boundary (CMB) condition are still missing. A recent experimental research concludes that water and silica have high mutual solubility at high pressure and high temperature, indicating that the boundary may be "fuzzy" [25]. However, the pressure range in their work is limited to about 100 GPa, which is far lower than the pressure range at core-mantle boundaries of Uranus and Neptune. Meanwhile, previous models of them suggest that the mantle is water-rich and also contains methane and ammonia [26,27] and both of them will decompose and release hydrogen under high pressures and temperatures [28–31]. In addition, carbon monoxide (CO) has been detected in their atmosphere, so a reaction $H_2O + CH_4 \rightleftharpoons CO + 3H_2$ is expected to take place in the interior [6,13]. Therefore, it is interesting to see whether $H_2O$ and $H_2$ can react with the core components.

In the deep interior of Uranus and Neptune, one of major components of the rocky core is silica [32–37], which attracts substantial interests in a lot of fields such as geophysics, warm-condensed matter physics and planetary sciences. Silica undergoes a series of phase transitions with compression [32–37], and recently a novel rhombohedral silica with mixed coordination is predicted to exist in the core of Neptune [37]. Several Si-O compounds with unusual stoichiometries have also been found [37]. Therefore, it is essential to consider the possibility of existence of silica-water/hydrogen compounds and their dynamical properties at the P-T conditions of Uranus and Neptune. To address these questions, we search for possible Si-O-H compounds in the pressure range up to 1 TPa based on crystal structure predictions and first-principles calculations. We have found three stable ternary compounds which may exist under the core-mantle boundary conditions of Uranus and Neptune. Furthermore, using *ab initio* molecular dynamics (AIMD), we revealed that the silica-water and the silica-hydrogen compounds exhibit superionic behavior, and the superionic region on the phase diagram covers the interior isentropic profiles of Uranus and Neptune, which indicate that they are competitive candidates for the conductivity source for the magnetic field of these giant planets.

Considering the predominant presence of Si, O and H in the core-mantle boundary, here we carried out variable-composition structure searching for Si-H, $SiO_2$-$H_2O$, $SiO_2$-H, and Si-O-H systems, from 300 GPa to 1 TPa. Interestingly, in the Si-O-H system, we identified three new ternary compounds which are thermodynamically stable under high pressures (Fig. 1(a)). They belong to the $SiO_2$-$H_2O$, $SiO_2$-H and $SiO_2$-$SiH_4$ pseudo-binary systems, respectively. The first ternary compound $Si_2O_5H_2$, satisfying the composition $(SiO_2)_2(H_2O)$, contains 36 atoms in the unit cell with space group of C2/c (Fig. 1(b)) and becomes stable above 450 GPa. According to the convex hull (Fig. 1(e)), the formation enthalpy with respect to silica and water is about 39 meV per formula at 450 GPa and 418 meV per formula at 700 GPa, indicating its energetic stability. All the hydrogen atoms in this structure are organized as bridges between two oxygen atoms, which is similar to some high-pressure phases of ice. At 450 GPa, the shortest O-H bond length is about 1.04 Å, which is very similar to that of ice (1.03 Å) calculated using the same parameters at the same pressure. Another compound, $SiO_2H_2$, is stable above 650 GPa. It has a silica-to-hydrogen ratio of 1:2 and 10 atoms in the unit cell, with space group of $P2_1/m$, as shown in Fig. 1(c). The formation enthalpy of $SiO_2H_2$ with respect to silica and hydrogen is about 41 meV per formula at 650 GPa and 64 meV per formula at 700 GPa (Fig. 1(f)). The third compound, $SiOH_2$, starts to be thermodynamic stable at around 500 GPa. It has 8 atoms in the unit cell and $P2_1/m$ symmetry (Fig. 1(d)), and its composition can be described as $Si(H_2O)$ or $(SiO_2)(SiH_4)$. In this $SiOH_2$ structure, the Si-O layers is separated by hydrogen atoms. Moreover, we also discovered another silica-water compound $SiO_3H_2$, with silica-to-water ratio of 1:1 and Pnma symmetry, is only 8 meV/atom above the convex hull at 650 GPa and becomes stable at around 800 GPa [38]. Silane [39,40] is found to decompose into SiH and $SiH_5$ above 700 GPa, as shown in Fig. 1(a). We have confirmed the phase diagram using the optB88-vdW functional [40]. The correction for the van der Waals interaction only slightly changes the stabilized pressures of compounds in the Si-O-H system, and does not affect the main conclusions. Moreover, the stable Si-O-H compounds are insulators with large band gaps. Details are shown in the supplementary material [38].

The electron localization function (ELF) of $Si_2O_5H_2$ show that there are two types of oxygen: the first one has covalent bonding with hydrogen atoms while the second one is isolated, as shown in Fig. 2(a). The interactions between silicon and oxygen atoms are ionic-like. These conclusions are further confirmed by the pCOHP between oxygen and hydrogen shown in Fig. 2(b). The atomic Bader charges are also calculated. Silicon and hydrogen lose 3.4 and 0.66 electron per atom respectively while every oxygen atom gets 1.6-1.7 electron. Considering the covalent bonds between oxygen and hydrogen, $Si_2O_5H_2$ is composed by three ions: $Si^{3.4+}$, $O^{1.7-}$ and $(HO_2)^{2.55-}$ with a ratio of 2:1:2. Detailed results about interactions analysis are shown in the supplementary material [38].

According to the current models of Uranus and Neptune [44], the conditions of their core-mantle boundaries reach pressures of 500-800 GPa and temperatures 5000-7000 K. To investigate the dynamical properties of the newly predicted compounds under these conditions, we carried out AIMD simulations within the pressure range 400-1100 GPa and the temperature range 2000-12000 K. Fig. 3 shows the mean squared displacements (MSD) and trajectories of hydrogen, oxygen, and silicon atoms in $Si_2O_5H_2$ at a density of 6.9 g/cm$^3$ and a temperature of 6000 K and in $SiO_2H_2$ at 794 GPa and 5000 K. The averaged pressure of the trajectory of $Si_2O_5H_2$ is about 485 GPa. This temperature-pressure condition is very close to core-mantle boundary of Uranus [44]. From these MSD curves (Fig. 3(a) and 3(b)), we found that the MSD value of hydrogen keeps increasing while MSD values of silicon and oxygen almost remain almost unchanged during the simulation. It means that protons diffuse freely like liquid while oxygen and silicon atoms only vibrate on their equilibrium sites. This is the feature of the superionic phase, which can also be confirmed by the trajectory (Fig. 3(c) and 3(d)).

The phase boundaries on the temperature-pressure phase diagram are identified by MSD, as shown in Fig. 4(a). Based on diffusive behaviors of the different elements, the phase diagram splits into three regions including solid, superionic and liquid phases. For comparison, we have also plotted the isentropes of Neptune and Uranus proposed by Nettelmann et al. [44] The phase boundary between the solid and superionic phases lies in the temperature range from 3000 to 4000 K with pressures from 400 to 900 GPa. $Si_2O_5H_2$ transforms from a superionic phase to a fluid at about 8000 K at 6.9 g/cm$^3$, with a pressure of 500 GPa. This high melting temperature is closed to that of silica under the same condition [2], indicating the stability of the Si-O framework. According to the P-T diagram, the region of superionic phase is broad enough to cover the core-mantle boundary conditions of Neptune and Uranus, indicating that superionic $Si_2O_5H_2$ is possible to be an important component at the deep mantle of these giants.

The silica-hydrogen compound $SiO_2H_2$ is also found to be superionic under high pressure and temperature according to MSD (Fig. 3(b)) and trajectories (Fig. 3(d)). The phase diagram of $SiO_2H_2$ is proposed in Fig. 4(b). The stabilized pressure of $SiO_2H_2$ is relatively higher so the compound is more likely to exist on Neptune rather than Uranus. In addition, we have plotted the P-T diagram for $SiOH_2$ in the supplementary material [38] but its superionic region is beneath the isentropes of Neptune and Uranus.

Moreover, due to its high melting temperature (11000K at 960 GPa), superionic $Si_2O_5H_2$ is possible to exist on the core-mantle boundary of Saturn (800-1800 GPa and 8000-12000 K) [45,46].

Another silica-water compound, Pnma $SiO_3H_2$ is also found to be superionic at high pressure and temperature (see supplementary material [38]). The stabilized pressure is too high for Uranus but it is possible to find superionic $SiO_3H_2$ on Neptune or larger giants such as Saturn [45,46]. Therefore, silica-water compounds found in this work may also be competitive candidate components of core-mantle boundary of Saturn.

The diffusion coefficient of hydrogen in superionic $Si_2O_5H_2$ is about $D_H = 4.8 \times 10^{-8}$ m$^2$/s at 485 GPa and 6000 K. We have calculated the electrical conductivity using the Nernst–Einstein equation: $\sigma = \frac{e^2 q_H^2 N_H D_H}{V k_B T}$, where $q_H = +1$ is the oxidation number of hydrogen, $N_H$ is the number of hydrogen atoms in the cell and $V$ is volume of the cell. The resulting electrical conductivity is around 8.9 S/cm.

The conductive shell responsible for non-dipolar and non-axisymmetric magnetic fields of Uranus and Neptune have been suggested to be metallic mixtures [49], metallic fluid hydrogen [50], superionic ice [9] and fluid trihydrogen oxide [24] proposed recently. These materials are expected to exist in ice-dominated environments, so ice giant models are favored in previous literature. The superionic silica-water and silica-hydrogen compounds found here are stable in the core-mantle boundary and core of giants. Their superionic states may be alternative important sources to explain the anomalous magnetic fields of Uranus and Neptune, which makes it possible to generate magnetic dynamos in the rock-dominated models. Based on our calculations and previous interior models [44], we proposed interior structures of Uranus and Neptune considering compounds predicted in this work, as shown in Fig.4(c). The silica-water compound is predicted to exist in the interior zone beneath about 7670 km of Uranus and about 9540 km of Neptune (white dashed lines). Our findings suggest that the core-mantle boundary of Uranus and Neptune might be "fuzzier" than we thought, because silica in the core can react with water and hydrogen in the deep mantle. Considering the high fraction of water in Uranus and Neptune, such reactions can play significant roles in the erosion of the core in these giants and may even have an impact on the choice between the compact and diluted/fuzzy cores [6,51].

In summary, we have predicted three Si-O-H compounds ($Si_2O_5H_2$, $SiO_2H_2$ and $SiOH_2$) which are stable at pressures corresponding to the deep mantle or core conditions of Uranus, Neptune and Saturn using crystal structure predictions and *ab initio* calculations. Furthermore, extensive *ab initio* molecular dynamics simulations demonstrate the existence of superionic phase of $Si_2O_5H_2$ at the core-mantle boundary conditions of Uranus, Neptune and Saturn. For another Si-O-H compounds $SiO_2H_2$, which belongs to the silica-hydrogen pseudo-binary system, the superionic region also covers the interior isentropic profiles of Neptune. These compounds resulted from unexpected chemical reactions under high pressure between very common natural materials - sand, water and hydrogen, suggest that they should be adopted into the models of giant planets. Importantly, the superionic states of these compounds also provide potential conductive candidates to explain the unusual magnetic fields in the rocky giant interior models. This may add scores to the rocky models rather than the current widely accepted icy models, and pave the way to ultimately solve the longstanding debate between the icy and rocky models. In addition, these compounds may also influence the demixing and erosion of the planet cores, which indicates that the transitional zone between the deep mantle and core of these giant planets may be gradual. These findings would be instructive to build more reliable models for Uranus and Neptune, which will enrich or even change the current knowledge of mankind about the giant planets in and outside the solar system.

**Acknowledgments**

J.S. acknowledges financial support gratefully from the National Natural Science Foundation of China (grant nos. 11974162 and 11834006), and the Fundamental Research Funds for the Central Universities. The calculations were carried out using supercomputers at the High Performance Computing Center of Collaborative Innovation Center of Advanced Microstructures, the high-performance supercomputing center of Nanjing University.

Figures:

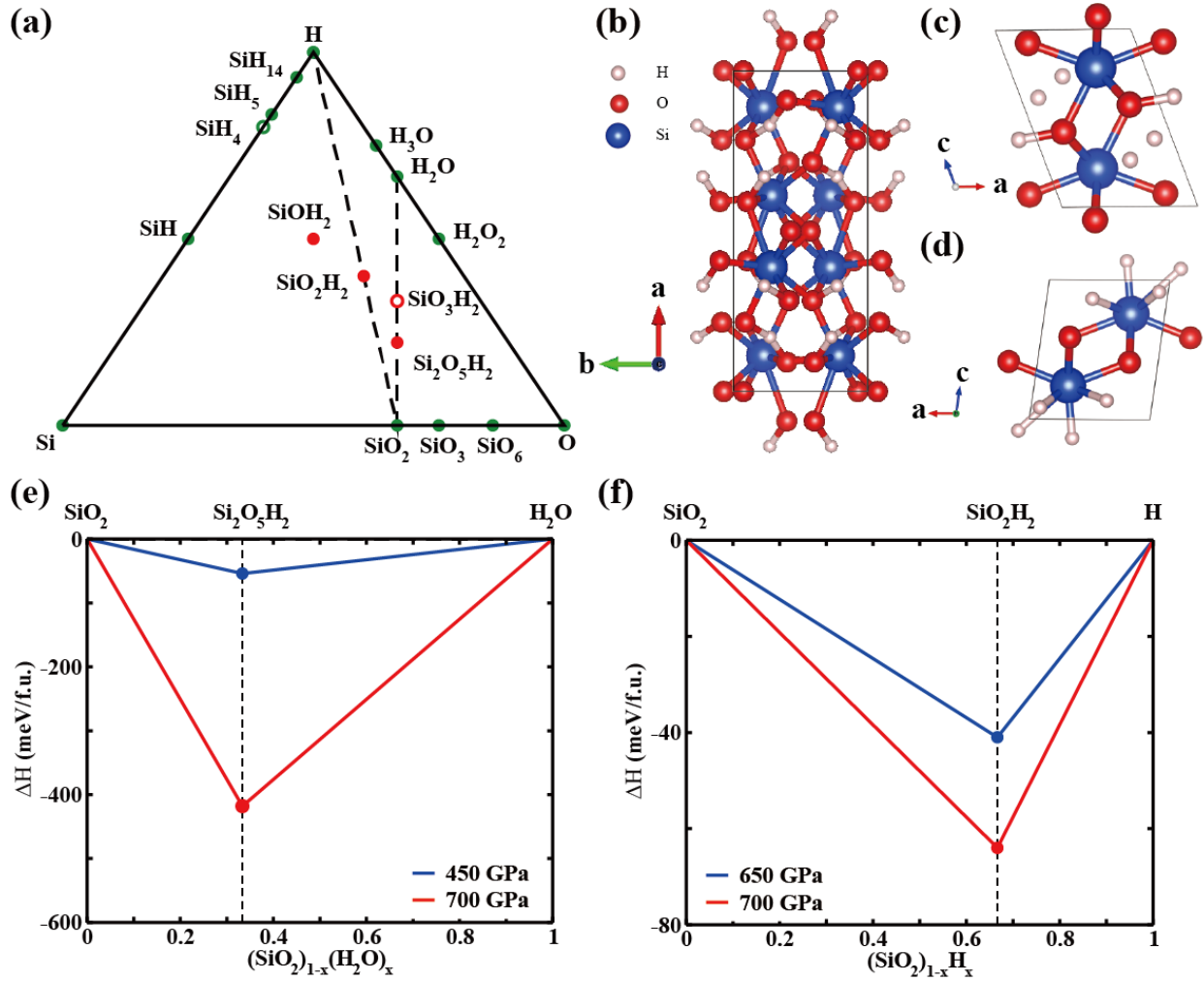

**Fig. 1. The energetics and crystal structures of the predicted Si-O-H compounds at high pressures.** (a) Si-O-H ternary phase diagram at 700 GPa. The known and newly found phases are marked by green and red, respectively. Solid and open circles represent the stable and metastable phases, respectively. The crystal structure of (b) the silica-water compound C2/c $Si_2O_5H_2$ (corresponding to $(SiO_2)_2(H_2O)$), (b) the silica-hydrogen compound $P2_1/m$ $SiO_2H_2$, and (d) the $P2_1/m$ $SiOH_2$. The convex hulls for (e) the $SiO_2$-$H_2O$ system and (f) the $SiO_2$-H system.

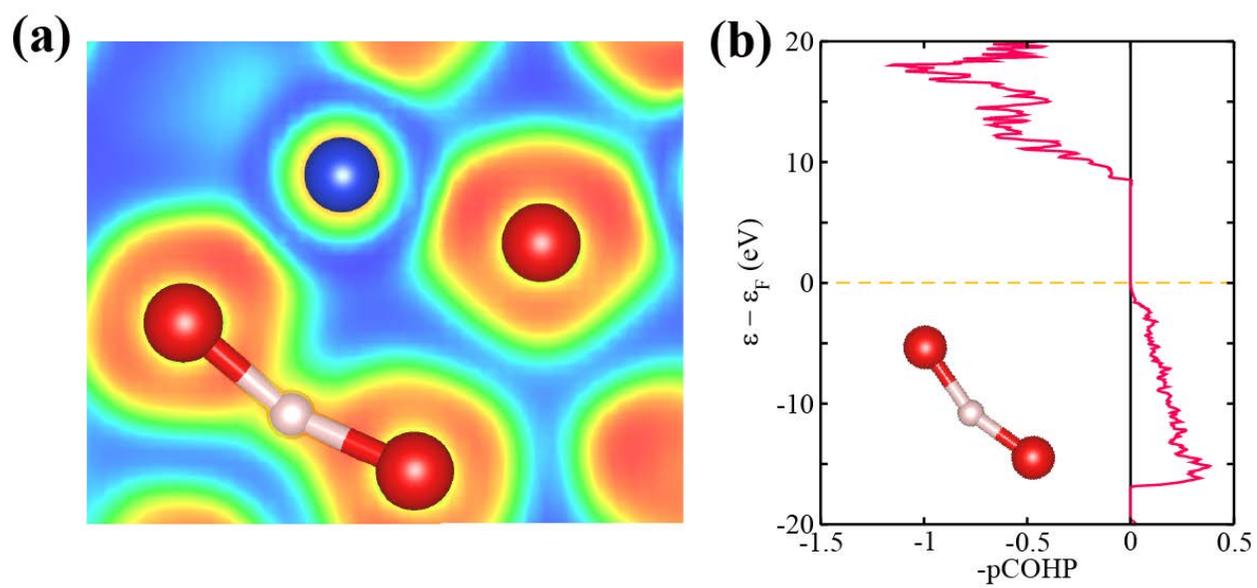

**Fig. 2. Electronic properties of the silica-water compound.** (a) ELF in a plane of $Si_2O_5H_2$ at 600 GPa. (b) The pCOHP analysis for O-H interactions in $Si_2O_5H_2$ at 600 GPa.

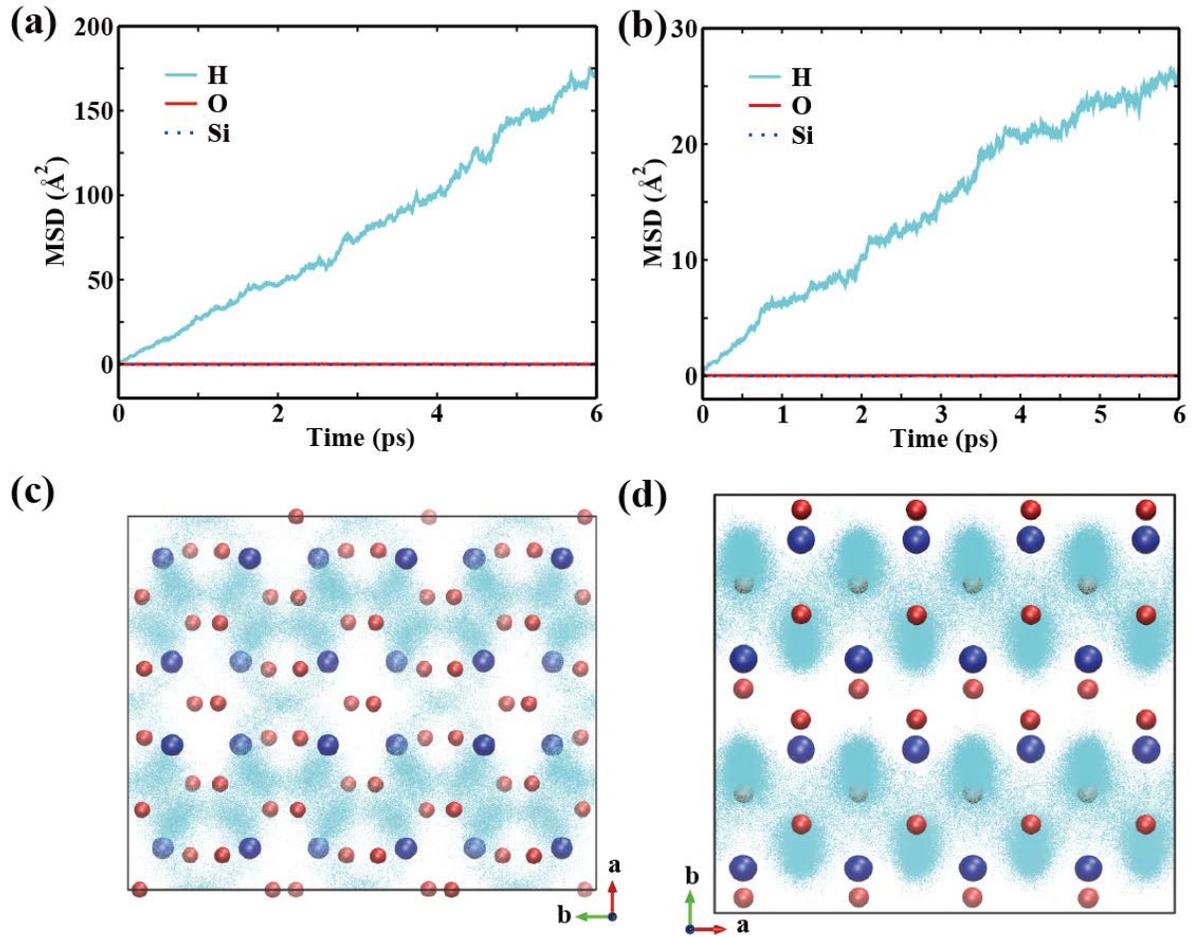

**Fig. 3. Dynamical properties of silica-water and silica-hydrogen compounds at high pressure and high temperature from ab initio molecular dynamics simulations.** MSD of hydrogen, oxygen and silicon in $Si_2O_5H_2$ (a) at 485 GPa and 6000K (corresponding to the isentropic condition of Uranus) and $SiO_2H_2$ (b) at 794 GPa and 5000 K (corresponding to the isentropic condition of Neptune). Trajectory snapshots of the corresponding superionic phase are shown in (c) and (d) respectively, where the blue and red balls represent equilibrium sites of silicon and oxygen atoms, respectively. The cyan points represent instantaneous positions of hydrogen atoms.

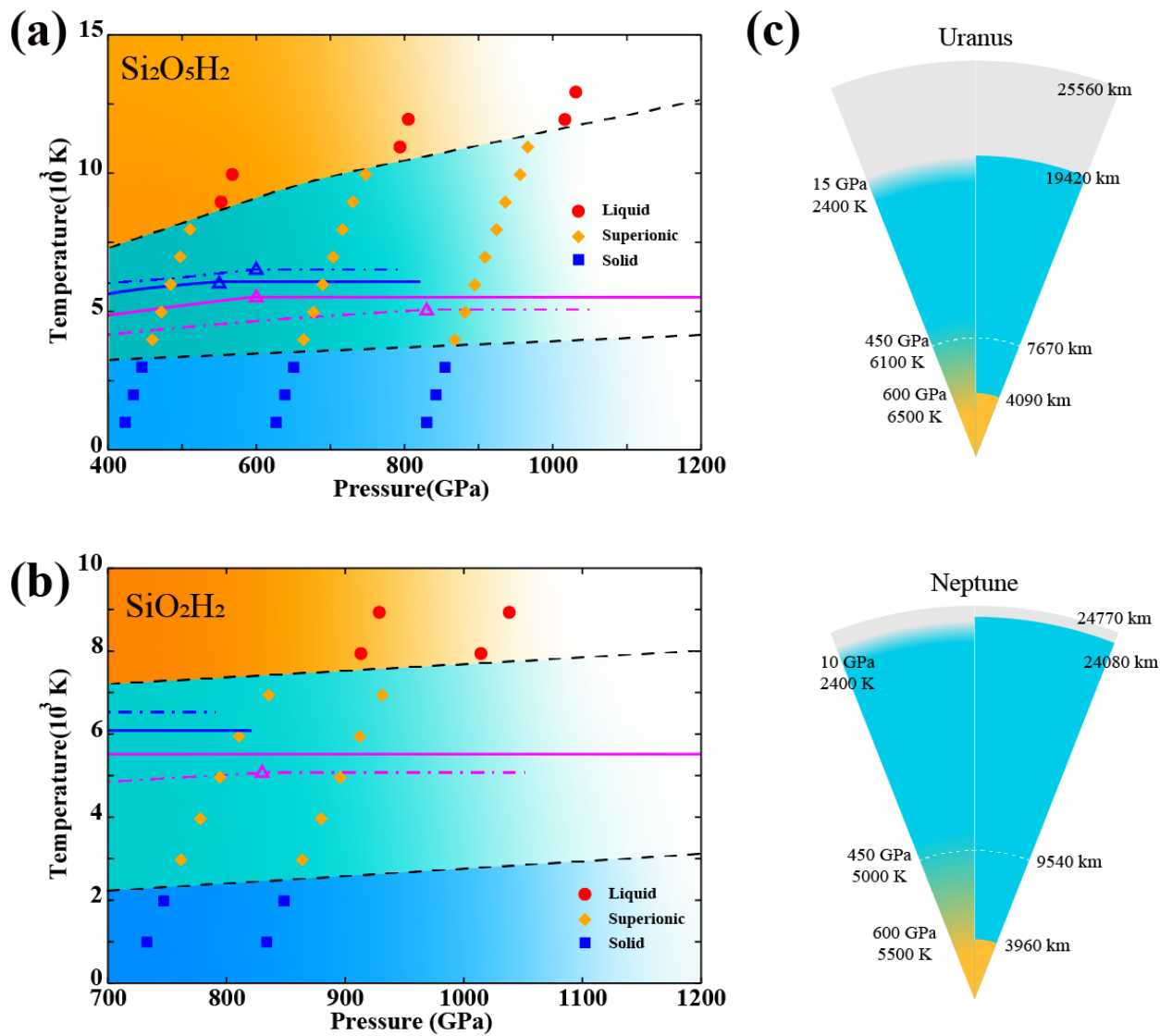

Fig. 4. Phase diagrams of silica-water and silica-hydrogen compounds and their impact on the interior structure of Uranus and Neptune. Phase diagrams of $Si_2O_5H_2$ (a) and $SiO_2H_2$ (b). Each solid symbol represents an AIMD simulation. The black dashed lines are phase boundaries. The blue and magenta lines represent interior isentropic profiles of Uranus and Neptune, respectively. For each planet, two different models from Ref. [44] marked by solid and dashed-dotted lines are presented. The open triangles are the states of core-mantle boundaries for each model. (c) Sketches of the internal structures of Uranus and Neptune based on the U2 and N1 models from Ref. [44]. The grey and cyan regions represent the H/He atmosphere and the "hot ice" layer composed by water/ammonia/methane, respectively, the yellow regions represent rocky cores. Both models with compact (right) and diluted/fuzzy (left) cores are shown. The cyan-yellow gradual change region beneath the white dashed curves represents the possible region for the silica-water compound predicted in this work.